\documentclass[preprint]{elsarticle}
\usepackage{lineno}
\usepackage[colorlinks,citecolor=blue,linkcolor=red,urlcolor=blue]{hyperref}
\usepackage{color}
\usepackage{fourier}
\usepackage{verbatim}
\usepackage{graphicx}
\usepackage{subfigure}
\usepackage{booktabs}
\usepackage{amsfonts}
\usepackage{multirow}
\usepackage{diagbox}
\usepackage{amsmath,amssymb}

\modulolinenumbers[5]

\begin{document}%\layout
%%%%%%%%%%%%%%%%%%%%%%%%%%%%%%%%%%%%%%%%%%%%%%%%%%%%%%%%%%%%%%%%%%%%%%%%%%%%%%%%%%%%%%%%%%%%%%%%
\begin{frontmatter}

\title{Favorable and unfavorable many-body interactions for near-field radiative heat transfer in nanoparticle networks}
%For at least  authors with different addresses, use instead the following commands

\author[mymainaddress,mysecondaddress]{Minggang Luo\corref{mycorrespondingauthor}}
\ead{minggang.luo@umontpellier.fr}

\author[mymainaddress,mythirdaddress]{Junming Zhao\corref{mycorrespondingauthor}}
\cortext[mycorrespondingauthor]{Corresponding author}
\ead{jmzhao@hit.edu.cn}

\author[myforthaddress]{Linhua Liu}

\author[mysecondaddress,myfifthaddress]{Mauro Antezza\corref{mycorrespondingauthor}}
\ead{mauro.antezza@umontpellier.fr}

\address[mymainaddress]{School of Energy Science and Engineering, Harbin Institute of Technology, 92 West Street, Harbin 150001, China}
\address[mysecondaddress]{Laboratoire Charles Coulomb (L2C) UMR 5221 CNRS-Universit\'e de Montpellier, F- 34095 Montpellier, France}
\address[mythirdaddress]{Key Laboratory of Aerospace Thermophysics, Ministry of Industry and Information Technology, Harbin 150001, China}
\address[myforthaddress]{School of Energy and Power Engineering, Shandong University, Qingdao 266237, China}
\address[myfifthaddress]{Institut Universitaire de France, 1 rue Descartes, F-75231 Paris Cedex 05, France}

\begin{abstract}
{Near-field radiative heat transfer (NFRHT) in nanoparticle networks is complicated due to the multiple scattering of thermally excited electromagnetic wave (namely, many-body interaction, MBI). The MBI regime is analyzed using the many-body radiative heat transfer theory at the particle scale for networks of a few nanoparticles. Effect of MBI on radiative heat diffusion in networks of a large number of nanoparticles is analyzed using the normal-diffusion radiative heat transfer theory at the continuum scale. An influencing factor $\psi$ is defined to numerically figure out the border of the different many-body interaction regimes. The whole space near the two nanoparticles can be divided into four zones, non-MBI zone, enhancement zone, inhibition zone and forbidden zone, respectively. Enhancement zone is relatively smaller than the inhibition zone, so many particles can lie in the inhibiting zone that the inhibition effect of many-body interaction on NFRHT in nanoparticle networks is common in literature. Analysis on the radiative thermal energy confirms that multiple scattering caused by the inserted scatter accounts for the enhancement and inhibition of NFRHT. By arranging the nanoparticle network in aspect of structures and optical properties, the MBI can be used to modulate radiative heat diffusion characterized by the radiative effective thermal conductivity ($k_{\rm eff}$) over a wide range, from inhibition (over 55\% reduction) to amplification (30 times of magnitude). To achieve a notable MBI, it is necessary to introduce particles that have resonances well-matched with those of the particles of interest, irrespective of their match with the Planckian window. This work may help for the understanding of the thermal radiation in nanoparticle networks.}
\end{abstract}


\end{frontmatter}

%\linenumbers

\section{Introduction}
Near-field radiative heat transfer (NFRHT) has attracted many attentions due to its rich physics. When the separation distance between two objects is comparable to or less than the thermal wavelength, near-field effects (e.g., evanescent wave tunneling effect) will dominate the radiative heat flux \cite{Rytov1989,Joulain2005,Shen2009}. The fluctuational electrodynamics theory proposed by Rytov et al. \cite{Rytov1989} was the basic theoretical framework to analyze NFRHT in different cases: two planar surfaces \cite{Loomis1994,Carminati1999,Shchegrov2000,Volokitin2001,Narayanaswamy2003,Volokitin2004} and two non-planar objects (two spheres \cite{Narayanaswamy2008}, one dipole and surface \cite{Chapuis2008plate} and two nanoparticles \cite{Chapuis2008,Manjavacas2012,Nikbakht2018}, respectively). Due to evanescent wave tunneling, the radiative heat flux between two objects (e.g., two plates \cite{Ottens2011,Lim2015,Watjen2016,Ghashami2018,DeSutter2019}, one plate and one sphere or tip \cite{Shen2009,Rousseau2009,Song2015} ) has been experimentally proved to be several orders of magnitude larger than the Plank's black-body limit. In the networks of many nanoparticles, nanoparticles often lie in the near field of each other, which leads to the significant multiple scattering effect of the thermally excited electromagnetic wave (namely, many-body interaction, MBI) affecting the NFRHT \cite{Tervo2017,Luo2020}.  Due to the complex many-body interaction, the approach to deal with NFRHT in two-body system cannot be directly applied in the system composed of many objects. To analyze NFRHT in the system composed of many particles, many theoretical frames at particle scale have developed, e.g., the many-body radiative heat transfer theory \cite{Ben2011,DongPrb2017}, trace formulas method \cite{Messina2011,Matthias2012,Messina2014,Muller2017} and the quasi-analytic solution \cite{Czapla2019}, etc. In addition, several important progress on NFRHT in dense particulate system at continuum scale has been reported \cite{Ben2013,Latella2018,Tervo2019,Ben2008,Tervo2016,Kathmann2018,Luo2021IJHMT_diff,Luo2023}.

For many-body interaction on NFRHT in the dense particulate system, some important progress has been reported recently. Ben-Abdallah \textit{et al.} \cite{Ben2011} proposed the many-body radiative heat transfer theory and revealed that the many-body interaction can significantly enhance the radiative heat flux between two SiC particles with insertion of a third particle at the center of the two particles for the first time. Such favorable many-body interaction for NFRHT is then investigated extensively for three-particle systems \cite{DongPrb2017,Wang2016AIP,Song2019IJHMT}. NFRHT between clusters (assemblies) composed of hundreds of dielectric SiC nanoparticles \cite{Dong2017JQ} (SiC-nGe core-shell nanoparticles \cite{Chen2018JQ}) is found to be inhibited significantly due to many-body interaction. In addition to the significant and obvious inhibition or enhancement on NFRHT, many-body interaction can also have weak and negligible effect on NFRHT between metallic
particle clusters at room temperature \cite{Luo2019}. Recently, many-body interaction on NFRHT in a chain of particle with another chain of particle in proximity is analyzed. The SiC proximate chain significantly inhibits the NFRHT in the chain of SiC particles due to coupling of localized surface phonon resonance (LSPhR). However, mismatch between characteristic thermal frequency at room temperature and polarizability resonance frequency of Ag nanoparitcle accounts for the weak many-body interaction on NFRHT in the chain of Ag nanoparticles with another Ag nanoparticle chain in proximity \cite{Luo2019JQ}. Moreover, the nearby substrate provides extra heat transfer channels significantly enhancing the radiative heat transfer between two particles or among a chain of particles \cite{DongPrb2018,Messina2018,Asheichyk2018,Zhang2019Prb,Yong2019prb_trans}. Furthermore, the geometric characteristics of the particulate systems themselves can also play a key role in NFRHT \cite{Phan2013,Nikbakht2017}. The findings reported in these works demonstrate that many-body interaction will have complex effect on radiative heat transfer.

The mechanism behind the many-body interaction on NFRHT in nanoparticle networks is still unclear. It is worthwhile to reveal how to distinguish the regime boarder of the enhancement and inhibition of many-body interaction on NFRHT. In order to understand insight of the regime boarder of many-body interaction on NFRHT, we extract two particles with a third inserted particle from a realistic three-dimensional many-particle system, as considered in the work by Tervo et al. \cite{Tervo2016,Tervo2017}. In this simplified three-particle system, due to insertion of a third scattering particle, we try to numerically figure out regime map of many-body interaction, i.e., judging whether and how the many-body interaction will enhance or inhibit NFRHT between two nanoparticles. The heat diffusion induced by the near-field radiative heat transfer in the many-body systems with a large number of particles (as opposed to just a few particles, like two- or three-particle system) can be characterized by the radiative effective thermal conductivity \cite{Ben2008,Tervo2016,Tervo2019,Kathmann2018,Luo2021IJHMT_diff,Luo2023}, which naturally includes the many-body interaction. It is still unclear how the many-body interaction affects the radiative heat diffusion in many-particle systems. Another missing point is whether we can take advantage of the MBI to manipulate radiative heat diffusion characteristics as desired. To simplify the problem, we extract a chain of large number of particles (typically hundreds of thousands of particles) with or without insertion of nearby identical particle chains out of the realistic three-dimensional many-particle system \cite{Tervo2016,Tervo2017}. In this simplified many-particle system, we try to numerically prove the possibility of controlling radiative heat diffusion characteristics by harnessing the many-body interaction.

This work is focused on effect of many-body interactions on near-field radiative heat transfer in nanoparticle networks by means of the many-body radiative heat transfer theory at particle scale \cite{Ben2011,DongPrb2017} and the normal-diffusion radiative heat transfer theory at continuum scale \cite{Luo2023}. Effect of insertion of a third particle (scatter) on NFRHT between two particles (emitter and receiver) is analyzed. Explicit regime boarder between enhancement and inhibition of many-body interaction on NFRHT is investigated in three-particle system. Feasibility of the manipulation of the effective thermal conductivity (radiative heat diffusion characteristics) by using the many-body interaction is discussed. 

\section{Theoretical aspect}
\label{Theorectical_aspect}

\begin{figure*}[htbp]
\centerline{\includegraphics[width=0.7\textwidth]{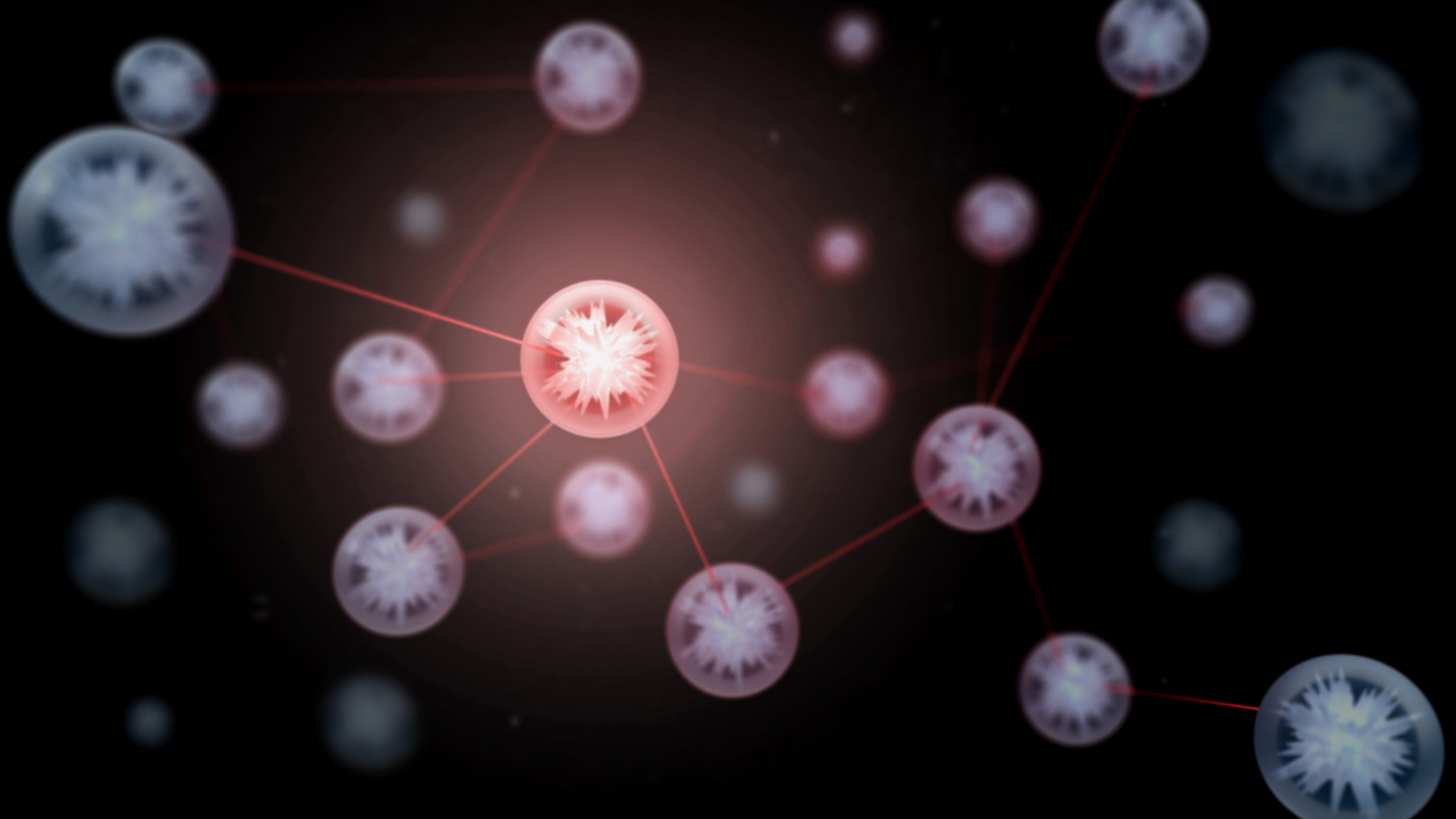}}
\caption{Structure diagram of the network of many particles. Heat diffusion induced by thermal radiation inside the nanoparticle network is investigated.}
\label{structure_MNPs}
\end{figure*}

%====== definition of keff / ETC for many-particle system =======%
For a many-particle system composed of $N_P$ small particles (particle radius is $a$), the radiative effective thermal conductivity is often used to characterize the heat diffusion induced by thermal radiation inside such systems (see Figure~\ref{structure_MNPs}), which naturally includes the many-body interactions.
%
%===== figure 0 Structure geometry=======%
According to our previous work on the normal-diffusion radiative heat transfer theory, the effective thermal conductivity tensor can be defined as \cite{Luo2023}
\begin{equation}
{\textbf{K}}_{\rm {eff}}=\frac{1}{2{V}_{\rm cell}} \sum_{i\ne P}^{{{N}_{P}}}{G_{iP} l_{i}^{2}{{\textbf{e}}_{i}}{{\textbf{e}}_{i}}},
\label{keff}
\end{equation}
where ${V}_{\rm cell}$ is the volume for the virtual cell occupied by the center particle $P$, $l_i=\vert \textbf{r}_i-\textbf{r}_P \vert$, $\textbf{r}_i$ and $\textbf{r}_P$ are positions of particle $i$ and particle $P$, $\textbf{e}_{i}= (\textbf{r}_i-\textbf{r}_P)/l_i$, $G_{iP}$ is the thermal conductance between particle $i$ and particle $P$, which is defined as \cite{Dong2017JQ,Luo2019}
\begin{equation}
G_{iP}^{}=\lim_{\delta T \rightarrow 0}\frac{\varphi_{j\leftrightarrow i}^{}}{T_P-T_i} ,
\label{Gij}
\end{equation}
where $\delta T=T_P-T_i$ is the temperature difference between the two particles, $\varphi_{P\leftrightarrow i}^{}$ is 
the net power exchanged between the $P$th and $i$th particles defined as
\begin{equation}
\begin{aligned}
\varphi_{P\leftrightarrow i}^{}=\varphi_{P\rightarrow i}^{}-\varphi_{i\rightarrow P}^{}=3\int_{0}^{+\infty} \frac{\mathrm{d}\omega}{2\pi}\left(\Theta(\omega,T_P)-\Theta(\omega,T_i)\right)\mathcal{T}_{i,P}(\omega),
\label{ExchangedPower}
\end{aligned}
\end{equation}
where $\omega$ is angular frequency, $\Theta(\omega,T)$ is the mean energy of the Planck oscillator, the power ($\varphi_{P\rightarrow i}^{}$) absorbed by the $i$th particle radiated by $P$th particle can be written as a Landauer-like formalism \cite{Ben2011,DongPrb2018}
\begin{equation}
\varphi_{P\rightarrow i}^{}=3\int_{0}^{+\infty} \frac{\mathrm{d}\omega}{2\pi}\Theta(\omega,T_P)\mathcal{T}_{i,P}(\omega) ,
\label{power}
\end{equation}
where the transmission coefficient $\mathcal{T}_{i,P}(\omega)$ between the $P$th and $i$th particles is \cite{Ben2011,DongPrb2017}
\begin{equation}
\mathcal{T}_{i,P}(\omega)=\frac{4}{3}k^4{\rm Im}\left(\chi_E^i\right){\rm Im}\left(\chi_E^P\right)\textrm{Tr}\left(G_{iP}^{EE}G_{iP}^{EE\dagger}\right),
\label{transmission}
\end{equation}
where $\chi_E^{}=\alpha_E^{}-\frac{ik^3}{6\pi}
\left|\alpha_E^{}\right|^2$, $\alpha_E$ is the electric polarizability, $k$ is vacuum wavevector, $G_{iP}^{EE}$ is the electric-electric Green’s function in the particulate system considering many-body interaction (The explicit expression can be found in Ref.~\cite{Luo2019}). For a small particle, the electric polarizability is $\alpha_E^{}=4\pi a^{3}(\varepsilon (\omega)-1)/(\varepsilon (\omega)+2)$ \cite{Chapuis2008}, where $\varepsilon (\omega)$ is the particle relative permittivity.

Since that the thermal radiation is spectrally dependent in general, the total radiative effective thermal conductivity tensor can be expressed as an integration of the spectral one among angular frequencies, ${\textbf{K}}_{\rm {eff}} =\int_0^{+\infty} {\textbf{K}}_{{\rm eff},\omega} \text{d}\omega$. For a symmetric and ordered many-particle system, the radiative effective thermal conductivity tensor reduces to a diagonal tensor with identical elements, $k_{\text{eff}}$. The radiative effective thermal conductivity of the network of many particles naturally includes the many-body interactions.

\begin{figure*}[htbp]
%\centerline{\includegraphics[width=0.7\textwidth]{structure.jpg}}
\centerline{\includegraphics[width=0.7\textwidth]{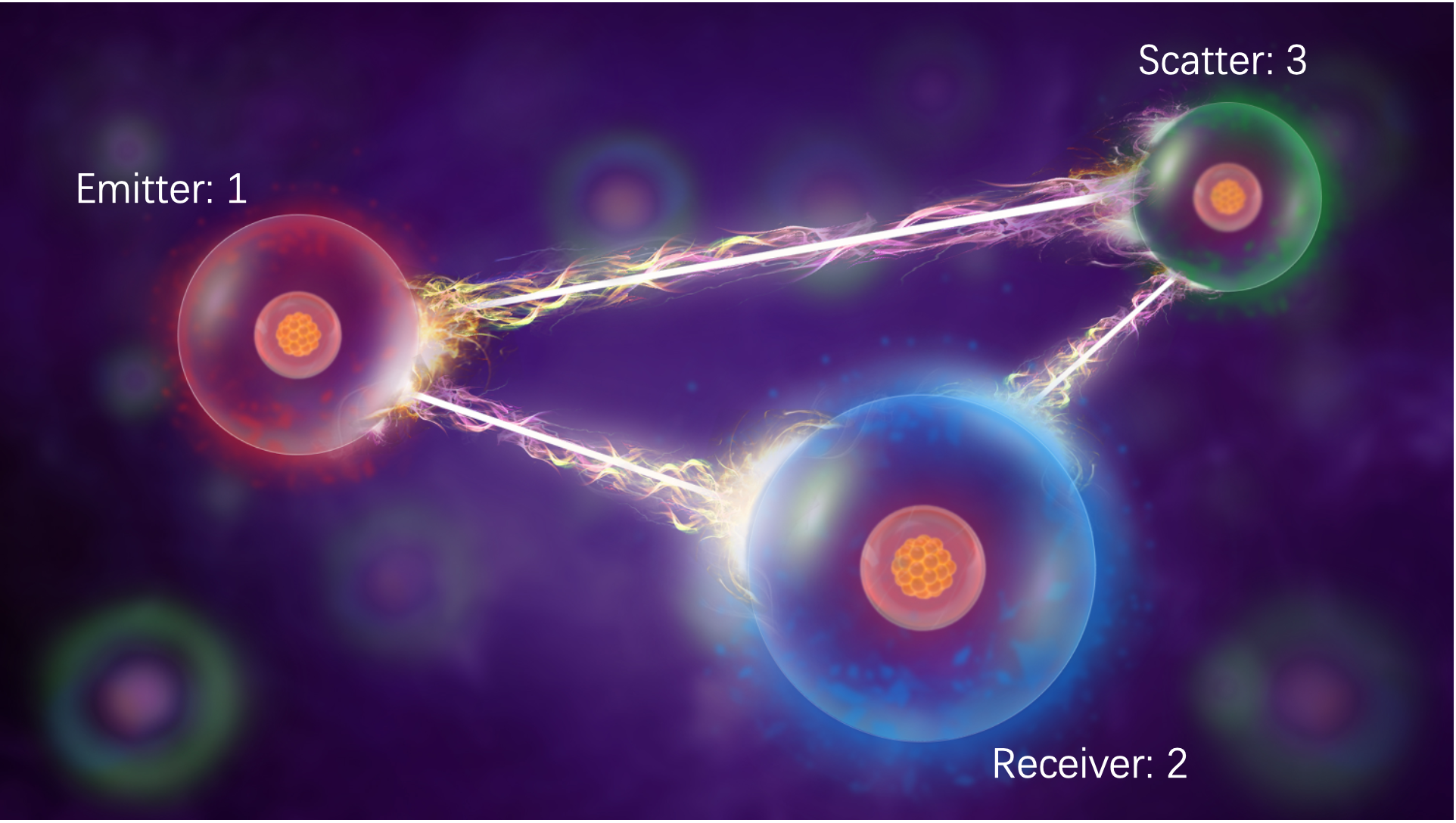}}
\caption{Structure diagram of the particulate system. NFRHT between emitter particle 1 (red) and receiver particle 2 (blue) with insertion of the scattering particle 3 (green) is investigated in this work.}
\label{structure}
\end{figure*}

To clearly understand the many-body interaction, we start from the network of only a few particles, of which the diagram is shown in Figure~\ref{structure}.
%===== figure 1 Structure geometry=======%
By using the many-body radiative heat transfer theory  at particle scale \cite{Ben2011, DongPrb2017}, the radiative heat flux and thermal conductance between the two particles separated by a certain separation distance $d$ are calculated. Considering the relative position of the scatter is strongly relevant to the radiative heat transfer between the emitter and receiver \cite{Ben2011, DongPrb2017}, we will give a clear regime map to explain the spatial dependence of scatter on radiative heat transfer. In addition, to understand insight of the different regimes of many-body interaction on radiative heat transfer, the thermally excited energy flux is investigated. The thermal energy flux at position $\textbf{r}$ radiated by the particle system can be evaluated by the Poynting vector, which is defined as \cite{Luo2020,Gchen2005book,Francoeur2008JQ}
\begin{equation}
\left \langle \textbf{S}(\textbf{r}) \right \rangle=2\int_{0}^{+\infty} \left \langle \textbf{S}(\textbf{r},\omega) \right \rangle\, \frac{\text{d}\omega}{2\pi} ,
\label{S}
\end{equation}
where the spectral Poynting vector $\left \langle \textbf{S}(\textbf{r},\omega) \right \rangle$ is
\begin{equation}
\begin{aligned}
\left \langle \textbf{S}(\textbf{r},\omega) \right \rangle=&2\sum_{i=1}^{N_P}\sum_{n=1}^{3}\sum_{m=1}^{3}{\rm Re} \left\{k^3\left[   \textbf{x}\left(G_{yn}^{EE}G_{zm}^{HE*}-G_{zn}^{EE}G_{ym}^{HE*}\right) \right. \right. \\&+\left.\left.\textbf{y}\left(G_{zn}^{EE}G_{xm}^{HE*}-G_{xn}^{EE}G_{zm}^{HE*}\right) +\textbf{z}\left(G_{xn}^{EE}G_{ym}^{HE*}-G_{yn}^{EE}G_{xm}^{HE*}\right)\right] {\rm Im}\left(\chi_E^{}\right)\Theta(\omega,T) \right\} ,
\label{S_spectral_final}
\end{aligned}
\end{equation}
where $m$ and $n$ are polarization direction index, $\textbf{x}$, $\textbf{y}$ and $\textbf{z}$ are the unit vectors of $x$, $y$ and $z$ axes in the given Cartesian coordinate system. $G_{\mu \tau}^{\nu E}$ ($\mu =x, y, z$; $\nu=E,H$ and $\tau=m,n=1,2,3$) is the element of the $3~\times~3$ electric-electric or magnetic-electric dyadic Green's function $G^{EE}$ or $G^{HE}$.

\section{Results and discussion}
Insertion of a third particle in the two-particle system can not only enhance \cite{Ben2011,DongPrb2017,Wang2016AIP,Song2019IJHMT} but also inhibit  \cite{Chen2018JQ,Luo2019,Luo2019JQ} the radiative heat flux between the two particles. However, it still remains unclear what the regime boarder is and how to clarify the regime boarder of many-body interaction on NFRHT, which is the focus of this work. NFRHT between two nanoparticles with insertion of a third particle is investigated in this work at room temperature. For networks of many nanoparticles (the number of particles is much large than 2 or 3), the effective thermal conductivity is commonly used to characterize heat diffusion induce by the NFRHT in such systems. The effect of many-body interaction on the effective thermal conductivity is analyzed. The nanoparticle radius ($a$) is 20 nm. The optical properties of the involved materials are: (1) The dielectric function of SiC is described by the Drude-Lorentz model $\varepsilon(\omega)=\varepsilon_{\infty}^{} (\omega^2-\omega_l^2+i\gamma\omega) / (\omega^2-\omega_t^2+i\gamma\omega)$ \cite{Palik}, where $\varepsilon_{\infty}^{}$ = 6.7, $\omega_l^{}$ = 1.827 $\times$ 10$^{14}$ rad$\cdot$s$^{-1}$, $\omega_t^{}$ = 1.495 $\times$ 10$^{14}$ rad$\cdot$s$^{-1}$, $\gamma$ = 0.9 $\times$ 10$^{12}$ rad$\cdot$s$^{-1}$, and (2) The dielectric functions of Ag is described by the Drude model $\epsilon(\omega) = 1-\omega_p^2/(\omega^2+i\gamma\omega)$ with parameters $\omega_p = 1.37 \times 10^{16}$ rad$\cdot$s$^{-1}$ and $\gamma$ = 2.732 $\times$ 10$^{13}$ rad$\cdot$s$^{-1}$ \cite{Ordal}. The separation distance between any two particles center to center is not less than 3$a$, which makes the dipole approximation valid \cite{DongPrb2017,SABEtAl2021}.

\subsection{MBI regime map for the NFRHT in networks of a few nanoparticles}
\begin{figure*} [htbp]
\centerline {\includegraphics[width=1.\textwidth]{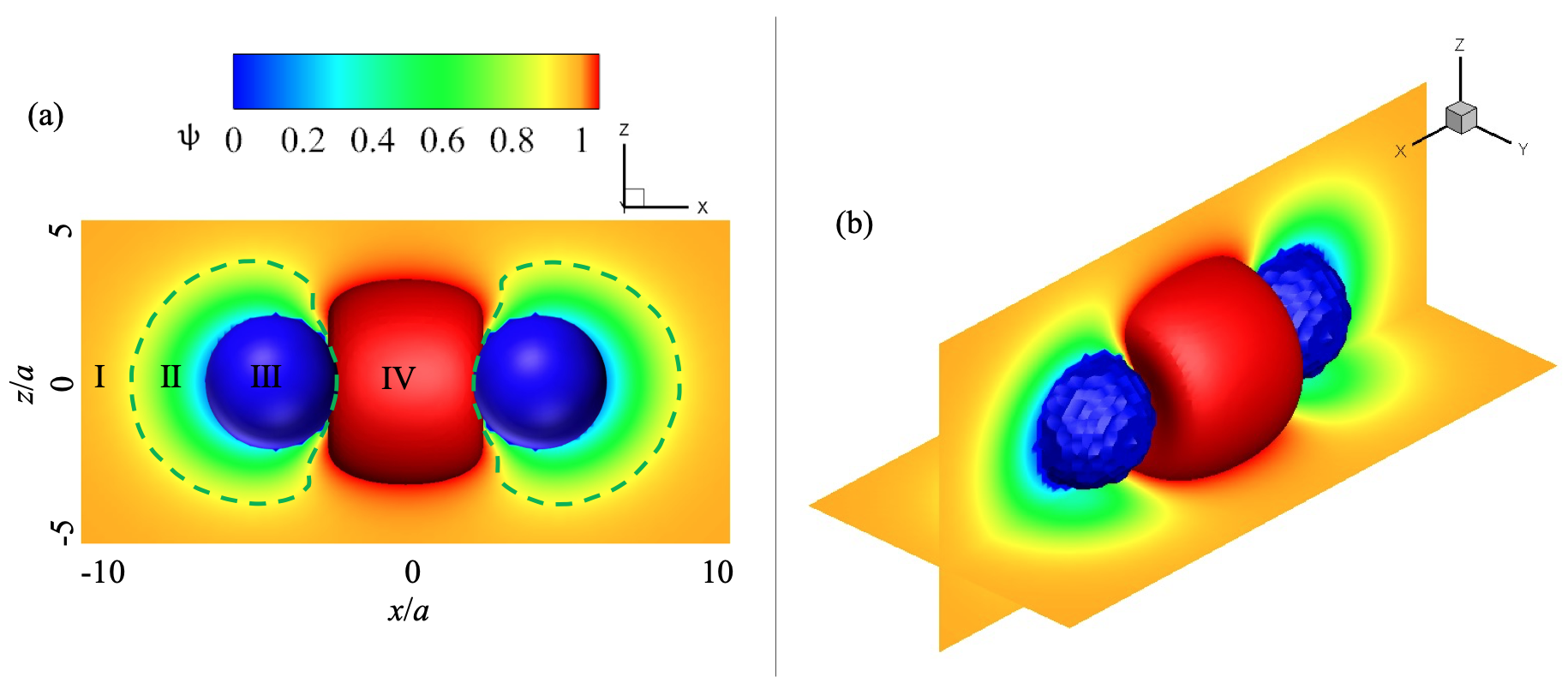}}
\caption{The dependence of the influencing factor $\psi$ on the position of the scatter from two different perspectives: (a) see from positive direction of $y$ axis to plane $xoz$; (b) an oblique perspective. The positions of the two nanoparticles are $(-4a,0,0)$ and $(4a,0,0)$.}
\label{phi}
\end{figure*}

The radiative heat flux between the emitter and receiver is dependent on the position of the inserted scattering particle. The dependence of the radiative thermal conductance $G$ between the emitter and the receiver on the position of the scatter in 3D domain is calculated. To quantitatively analyze the effect of many-body interaction on NFRHT due to insertion of a third particle, an influencing factor $\psi$ is defined as 
\begin{equation}
\psi=\frac{G}{G_0},
\label{IF}
\end{equation}
where $G_0$ is the thermal conductance between the emitter and receiver solely without insertion of a third particle. $\psi$ is dependent on the position of the inserted scattering particle. The dependence of the influencing factor $\psi$ on the position of the scatter is shown in Figure~\ref{phi} for a three-particle system (we take the SiC nanoparticles as an example). The separation $d$ between the emitter particle and receiver particle is $8a$. The emitter and receiver are near to thermal equilibrium around 300 K. The scattering particle is fixed at 0 K to study the radiative heat transfer purely between the emitter and receiver.

%===== figure 2 regime maps=======%

As shown in Figure~\ref{phi}, the whole space around the two nanoparticles can be divided into four zones, i.e., \emph{Zone I} (non-MBI zone, $\psi \approx 1$); \emph{Zone II} (inhibition zone, $0<\psi<1$); \emph{Zone III} (forbidden zone, $\psi \equiv 0$) and \emph{Zone IV} (enhancement zone, $\psi>1$). \emph{Zone I}: $G\sim G_0$. Insertion of a third nanoparticle in this zone has a negligible effect on the thermal conductance between two nanoparticles. The influencing factor $\psi\approx 1$. In this zone, the third nanoparticle lies far away from the two nanoparticles. \emph{Zone II}: $G<G_0$. $0<\psi<1$. When the inserted nanoparticle lies near the emitter or absorber nanoparticles, the thermal conductance is significantly inhibited. \emph{Zone III}: forbidden zone in blue $(\psi \equiv 0)$, which is defined as $|\textbf{r}-\textbf{r}_{\nu}|<2a,~(\nu=1,2)$, where the $\textbf{r}$ is the position vector of an arbitrary point and $\textbf{r}_{\nu}$ is the position vector of center of the nanoparticle 1 and 2 ($\nu=1,2$). No nanoparticle can be inserted in this forbidden zone. Because no overlap of nanoparticles is allowed. \emph{Zone IV}: $G>G_0$. The enhancement zone (red zone in Figure~\ref{phi}) is defined by $\psi>1$, where insertion of a third nanoparticle is in favor of NFRHT between the two nanoparticles. The inserted nanoparticle in the enhancement zone can be treated as the intermediate used to enhance NFRHT. More thermal energy is exchanged between the two nanoparticles due to coupling between this intermediate nanoparticle and the two nanoparticles. Due to the multiple scattering of the excited thermal wave by the inserted nanoparticle, the radiative thermal energy absorbed by the nearby nanoparticle is decreased slightly, which accounts for the inhibition of NFRHT caused by the insertion of a third nanoparticle very close to the two particles. In addition, the enhancement zone IV is relatively smaller than the inhibiting zone II. For any two nanoparticles in the realistic particulate system composed of hundreds of thousands of nanoparticles, it is inevitable that there are many nanoparticles lying in the inhibiting zone, which accounts for the inhibition on NFRHT in many-particle system observed in Refs.~\cite{Dong2017JQ,Chen2018JQ,Luo2019JQ,Luo2019}.

\begin{figure} [htbp]
\centerline {\includegraphics[width=0.4\textwidth]{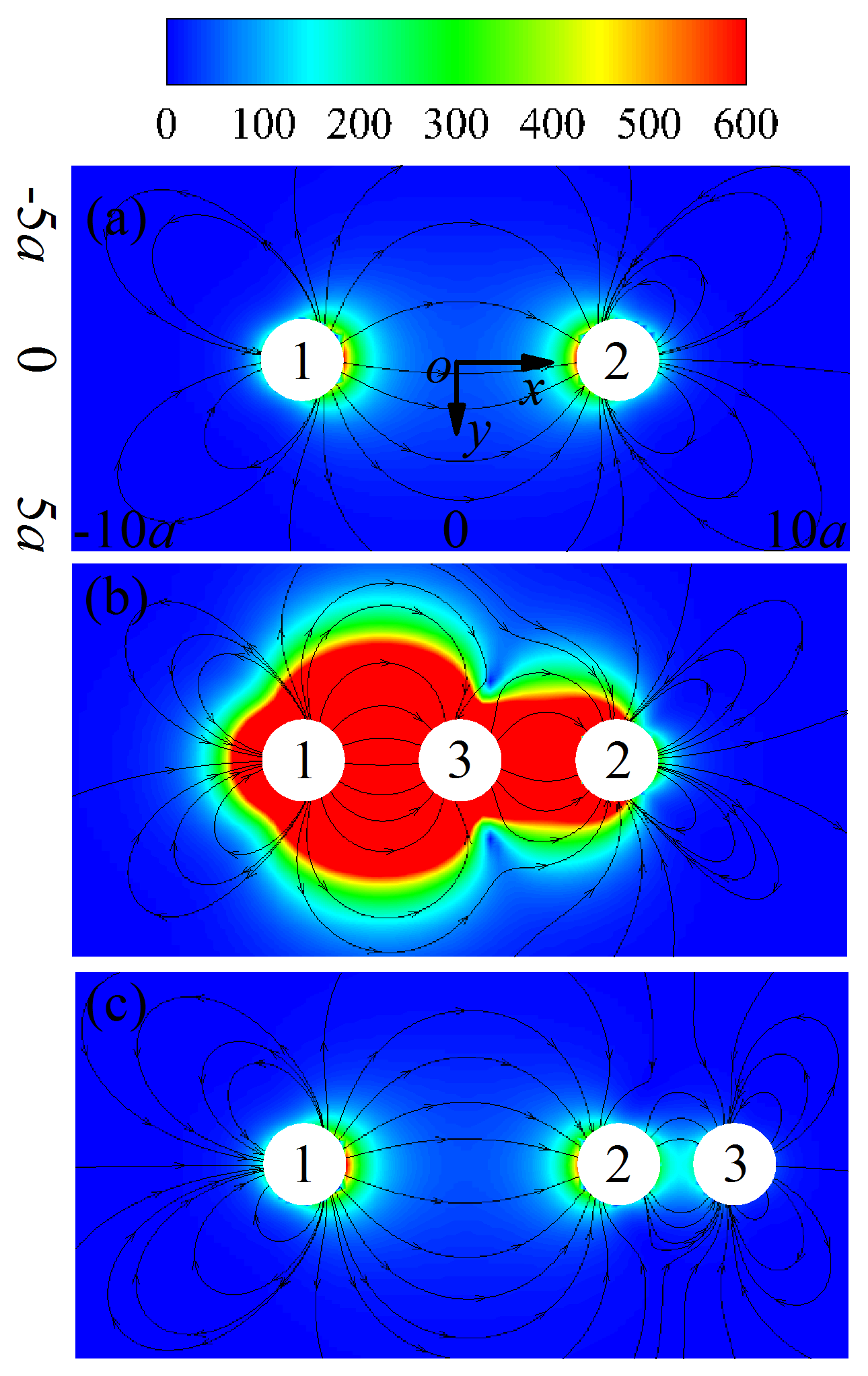}}
\caption{Magnitude of Poynting vector $\textbf{S}$ in the plane $xoy$ for three cases: (a) 2 nanoparticles without insertion of a third nanoparticle; (b) the third particle is inserted in the central point of the connection line between the two nanoparticles, and (c) the third particle is inserted in the proximate point of one particle in extension line of the connection line between the two particles.}
\label{S_total}
\end{figure}

%\subsection{Radiative thermal energy in networks of a few nanoparticles}
From above analysis, insertion of a third nanoparticle in the enhancement zone between two nanoparticles will enhance NFRHT, while insertion of a third nanoparticle in the inhibition zone between two nanoparticles will inhibit NFRHT. To understand insight of these two different regime of many-body interaction on NFRHT, Poynting vector thermally excited is investigated. Magnitude of Poynting vector $\textbf{S}$ in the plane $xoy$ for three cases are shown in Figure~\ref{S_total}: (a) 2 nanoparticles without insertion of a third nanoparticle, (b) the third scattering particle is inserted at the center of the connection line between the emitter and receiver, and (c) the third scattering particle is inserted in the proximate point of receiver particle in extension line of the connection line between the two particles, respectively. Thermal energy flux streamlines are also added for reference. The size of the calculation domain is $20a~\times~10a$. The white zones are for the nanoparticles. The coordinates of particle 1 and 2 are $(-4a,0,0)$ and $(4a,0,0)$. For Figure~\ref{S_total}(b) and (c), coordinate of particle 3 is $(0,0,0)$ and $(7a,0,0)$, respectively. The emitter particle 1 is fixed at 300 K. The receiver particle 2 and the scattering particle 3 are fixed at 0 K to study the emission purely given by the emitter.

%===== figure 3 Poyting vector contour=======%

As can be seen in Figure~\ref{S_total}(b), the Poynting vector is much larger than that observed in Figure~\ref{S_total}(a) and (c). From the thermal energy flux streamlines in Figure~\ref{S_total}, the inserted scattering particle in the center of the connection line between the emitter and receiver works as a relay (an intermediate) of radiative thermal energy. The significant coupling in this case (b) accounts for the enhancement of radiative thermal energy, hence more energy can be emitted by the emitter. As shown in Figure~\ref{S_total} (a) and (c), the Poynting vector distribution for these two cases are similar to each other. Insertion of a third scattering particle in proximate point of receiver particle in the extension line of the connection line between the two particles has a weak inhibitive effect on radiative thermal energy, which is corresponding with the weak inhibition zone II observed in Figure~\ref{phi}. However, from the Figure~\ref{S_total} (a) and (b), insertion of a third scattering particle in the center of the connection line between the two particles significantly enhances the radiative thermal energy, which is corresponding with the enhancement zone IV observed in Figure~\ref{phi}.

\subsection{Heat diffusion characteristics induced by thermal radiation in networks of many nanoparticles}

In this section, effect of many-body interactions on the heat diffusion characteristics induced by the thermal radiation in networks of many nanoparticles is analyzed. We use the effective thermal conductivity (ETC, $k_{\rm eff}$) to characterize the effective heat diffusion induced by the NFRHT in networks of 1000 nanoparticles, which is generally sufficient to get convergent results \cite{Luo2021IJHMT_diff,Luo2023}. According to previous analysis, if we put the particles in the enhancement \textit{zone IV} (i.e., the center of two particles), there will a significant enhancement on the thermal radiation, which is expected to also enhance the radiative effective thermal conductivity $k_{\rm eff}$. However, if we put the particles in the inhibition \textit{zone II} rather than the enhancement \textit{zone IV}, the NFRHT will be significantly inhibited rather than enhanced, which is thus expected to inhibit the $k_{\rm eff}$ consequently. It will be worthwhile to confirm whether the heat diffusion characteristics of nanoparticle networks can be manipulated by means of many-body interactions.

\begin{figure} [htbp]
\centerline {\includegraphics[scale=0.4]{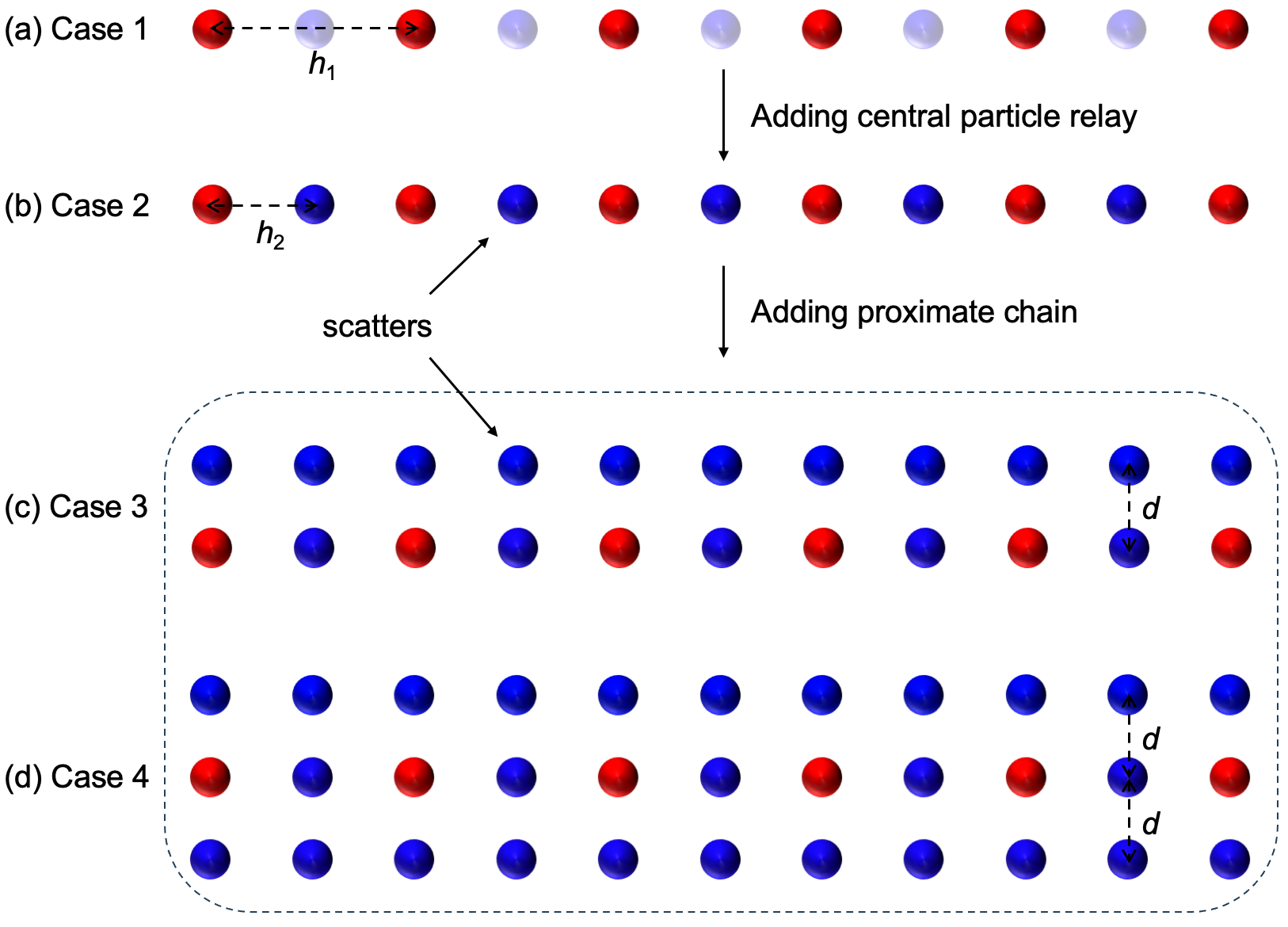}}
\caption{Diagram of the geometry for the considered four cases: (a) case 1 (period of the structured nanoparticle chain is $h=6a$); (b) case 2 (adding a particle relay in arbitrary two adjacent particles in case 1, then the period is $h=3a$.); (c) case 3 (adding a proximate chain in case 2); (c) case 4 (adding two proximate chains in case 2). The minimum separation between two particles in these cases is 3$a$, which still guarantees the validity of the dipolar approximation \cite{DongPrb2017,SABEtAl2021}.}
\label{ETC_geo}
\end{figure}

We consider four numerical cases to analyze effect of MBI on the radiative heat diffusion characteristics. The geometries of the four cases are shown in Figure ~\ref{ETC_geo}. In the Case 1, the period of the structured and ordered particle chain is $h_{\rm 1}=6a$. As for the Case 2, we add a particle relay in arbitrary two adjacent particles in case 1, of which the positions are shown as shading of Case 1 of Figure ~\ref{ETC_geo}. Thus the new particle chain in the Case 2 is a new structured and ordered particle chain with a new period $h_{\rm 2}=3a$. Then by adding one proximate chain or two proximate chains to the chain in Case 2 with a separation $d=3a$ between the two chains, we have Case 3 and Case 4, respectively.

%===== figure 4 geometry scheme======%

\begin{figure*} [htbp]
\centerline {\includegraphics[width=1.\textwidth]{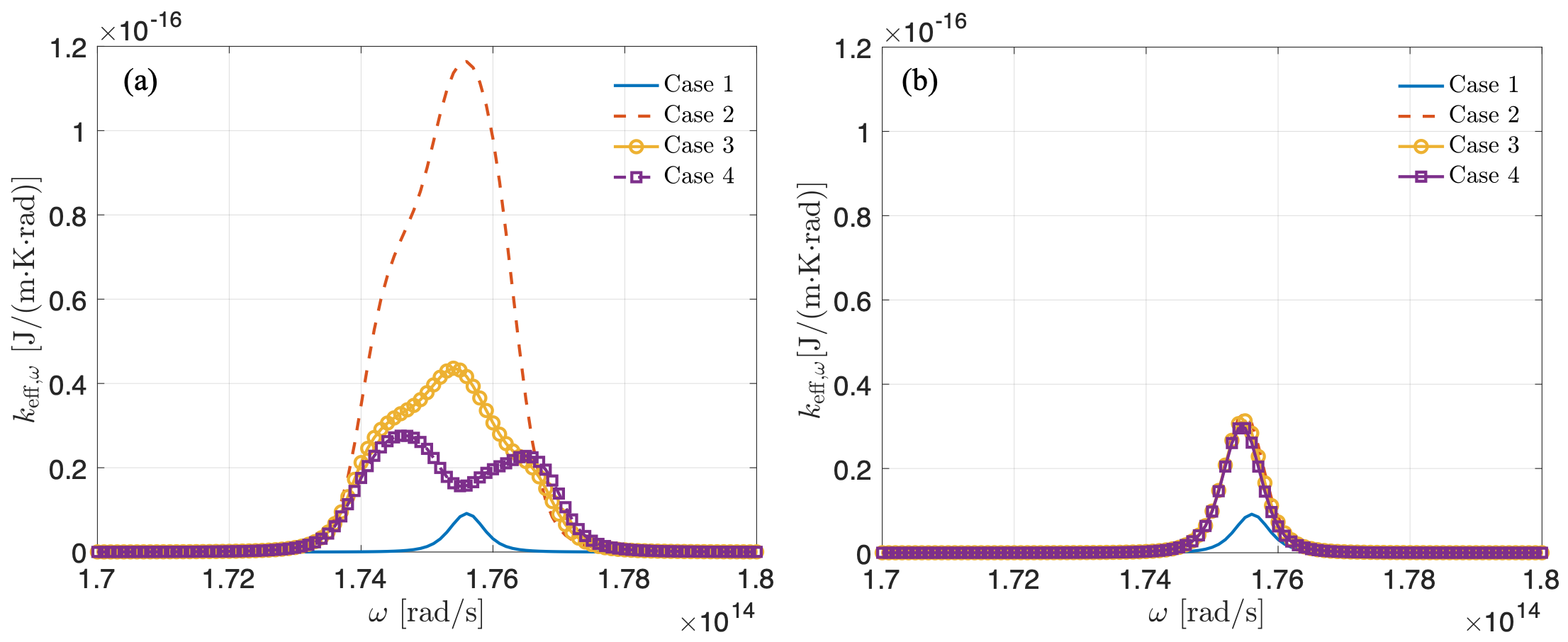}}
\caption{Spectral effective thermal conductivity for the four considered cases: (a) SiC scatters, and (b) Ag scatters}
\label{ETC_spectrum}
\end{figure*}

The radiative effective thermal conductivity $k_{\rm eff}$ is calculated at 300 K by using the Eq.~(\ref{keff}). The particle chain in Case 1 is composed of dielectric SiC. We start the analysis from the SiC scatters, (i.e., the blue scatters in Case 2 to Case 4 are all considered as the SiC particles). The total radiative thermal conductivity of the four cases from Case 1 to 4 are 6.77 $\mu$W/(m$\cdot$K), 230.84 $\mu$W/(m$\cdot$K), 102.10 $\mu$W/(m$\cdot$K), and 75.45 $\mu$W/(m$\cdot$K), respectively. The corresponding spectrum of the effective thermal conductivity for the four cases is shown in Figure ~\ref{ETC_spectrum} (a). By comparing the total effective thermal conductivity of the Case 1 and Case 2, the MBI significantly enhances the radiative heat diffusion in the nanoparticle networks by about 34 times of magnitude. As shown in Figure ~\ref{ETC_spectrum} (a), the resonance peak of the spectrum for Case 2 improves a lot as compared to Case 1, when inserting particles in the enhancement zone. The added particles in the enhancement zone work as relays for near-field photon tunneling and thus greatly enhance radiative heat exchange. Moreover, by comparing the total effective thermal conductivity of the Case 2 and Case 3, the MBI significantly reduces the radiative effective thermal conductivity by about 55\%, when inserting proximate particles in the inhibition zone.
By adding one more proximate chain (Case 4) in the Case 3, the effective thermal conductivity further reduces by about 26\%. Through specific structure arrangement of the particles in the network, the MBI can be applied to manipulate radiative heat diffusion as desired with a high modulation ratio, ranging from inhibition to amplification.

%===== figure 5 spectrum of ETC for C1-C4=======%

\begin{figure*} [htbp]
\centerline {\includegraphics[width=0.65\textwidth]{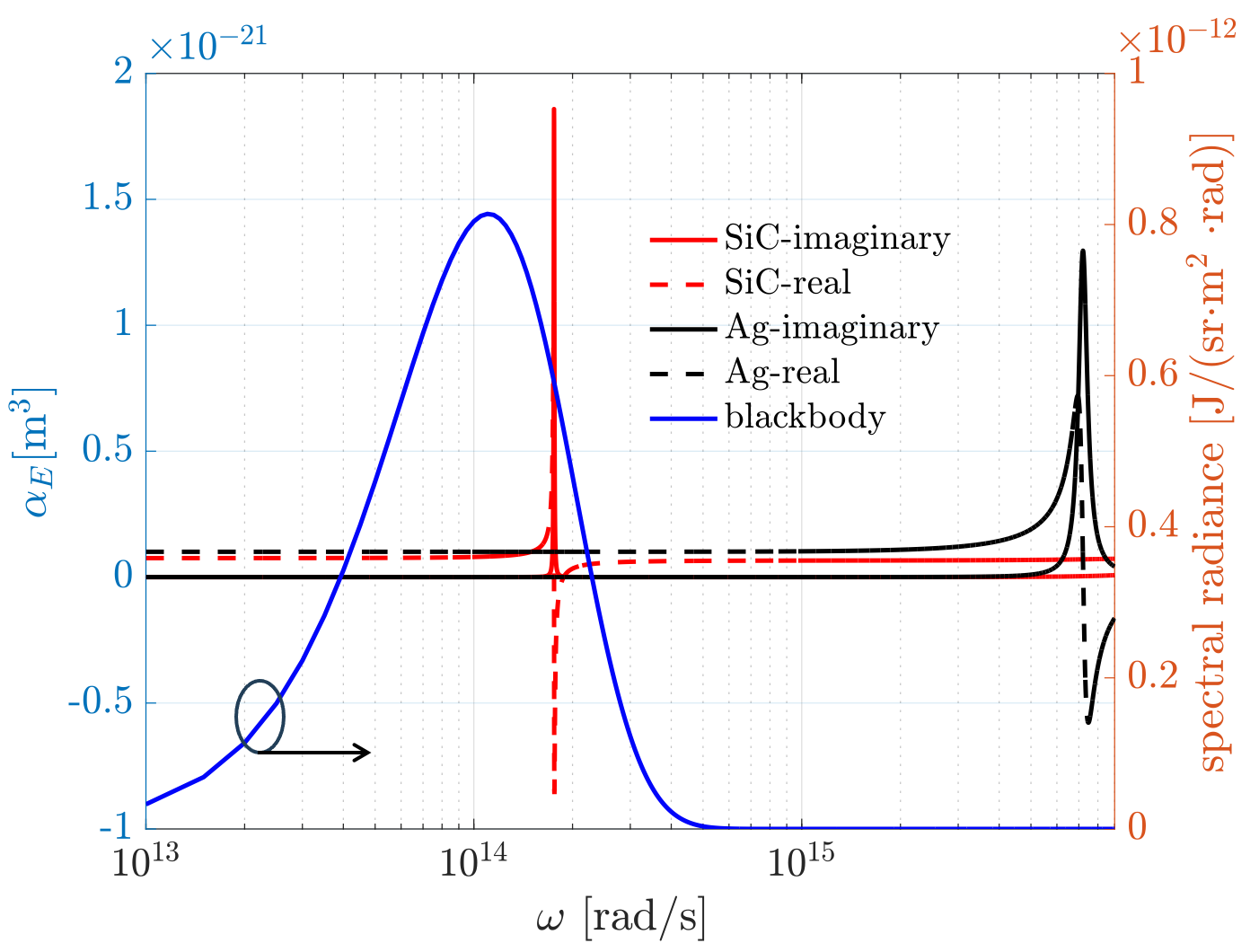}}
\caption{Polarizability spectrum of the Ag and SiC particles (radius $a=20$nm). The spectral radiance of the blackbody at 300 K is added for reference.}
\label{Ag_SiC_polarizibility}
\end{figure*}

We then consider the SiC chain but with Ag scatters (i.e., the blue scatters in Case 2 to Case 4 in Figure ~\ref{ETC_geo} are all considered as the Ag particles). The total radiative thermal conductivity of the four cases from Case 1 to 4 are 6.77 $\mu$W/(m$\cdot$K), 27.74 $\mu$W/(m$\cdot$K), 25.66 $\mu$W/(m$\cdot$K), and 24.41 $\mu$W/(m$\cdot$K), respectively. The corresponding spectrum of the effective thermal conductivity for the four cases is shown in Figure ~\ref{ETC_spectrum} (b). By comparing the total effective thermal conductivity of the Case 1 and Case 2, the MBI significantly enhances the radiative heat diffusion in the nanoparticle networks by about 4 times of magnitude. As shown in Figure ~\ref{ETC_spectrum} (b), the resonance peak of the spectrum for Case 2 improves a little as compared to Case 1, when inserting Ag scattering particles in the enhancement zone. Although the added Ag particles in the enhancement zone also can work as relays for near-field photon tunneling and thus greatly enhance radiative heat exchange, the enhancement to the effective thermal conductivity induced by the Ag scatter relays (4 times enhancement) is much weaker than that of the SiC scatter relays (34 times enhancement). Moreover, by comparing the total effective thermal conductivity of the Case 2 and Case 3, the MBI significantly reduces the radiative effective thermal conductivity by about 7.5\%, when inserting proximate Ag scattering particles in the inhibition zone.
By adding one more proximate Ag scattering chain (Case 4) in the Case 3, the effective thermal conductivity further reduces by about 4.9\%. In general, that is to say the Ag scattering particles can bring a relatively weak effect on the radiative effective thermal conductivity than the SiC scattering particles do. Because the coupling in the SiC chain with SiC proximate scattering chain is much stronger than that in the SiC main chain with Ag proximate scattering chain. As we can see from the polarizability spectrum of the SiC and Ag nanoparticles in Figure ~\ref{Ag_SiC_polarizibility} (the blackbody's spectral radiance at 300 K is also added for reference), there is an obvious mismatch between the localized surface resonances supported by the two types of nanoparticles (the former in the Planckian window, the latter at the optical frequency).

%===== figure 6 polarizability spectrum of Ag and SiC=======%

% For a SiC proximate chain with a fixed the separation $d$, as shown in Figure~\ref{Ag_SiC_proxi_MBI}, the ratio $k_{\rm eff}/k_{\rm eff,0}$ obviously increases with the period $h$, which is due to the decreasing coupling along the chain. However, for the Ag proximate chain, the $k_{\rm eff}/k_{\rm eff,0}$ does not change too much when changing the period of the main chain, due to the weak coupling in such system mediated by the mismatch between localized surface resonances of Ag and SiC nanoparticles.

\begin{figure*} [htbp]
\centerline {\includegraphics[width=1.\textwidth]{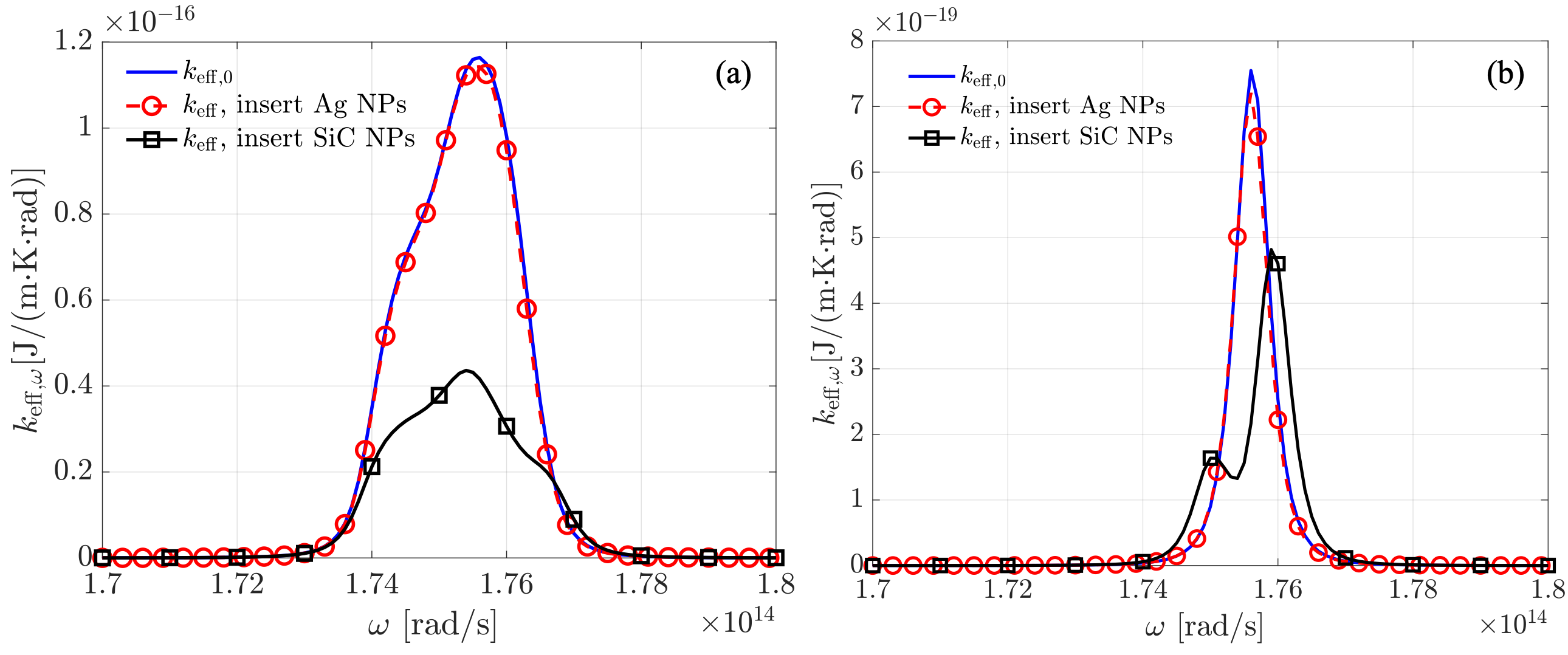}}
\caption{The effective thermal conductivity spectrum of the main chain (SiC particles) with and without scatters: (a) $h=3a$, and (b) $h=10a$.}
\label{proxi_Ag_SiC_spectrum}
\end{figure*}

We take the cases of period $h=3a$ (60 nm) and $10a$ (200nm) for instance to show the spectrum of effective thermal conductivity in Figure~\ref{proxi_Ag_SiC_spectrum}. For the two considered periods, as shown in Figure~\ref{proxi_Ag_SiC_spectrum} (a) and (b), when having Ag scattering proximate chain, the effective thermal conductivity spectrum is nearly identical to that of the SiC main chain without any additional scatters. As mentioned before, there is big mismatch between the localized surface resonances of Ag and SiC nanoparticles. Therefore, the coupling in the SiC main chain itself is not affected, regardless of whether the Ag proximate scatters are present or not. However, when having a SiC scattering proximate chain, the peak of the spectrum drops dramatically from the blue line to the black line for both periods considered, because more and more SiC particles enter the inhibition zone and thus the effective thermal conductivity decreases significantly.

%===== figure 8 spectrum of ETC=======%

%======== effect of separation between main chain and proximate chain on keff

\begin{figure*} [htbp]
\centerline {\includegraphics[width=0.7\textwidth]{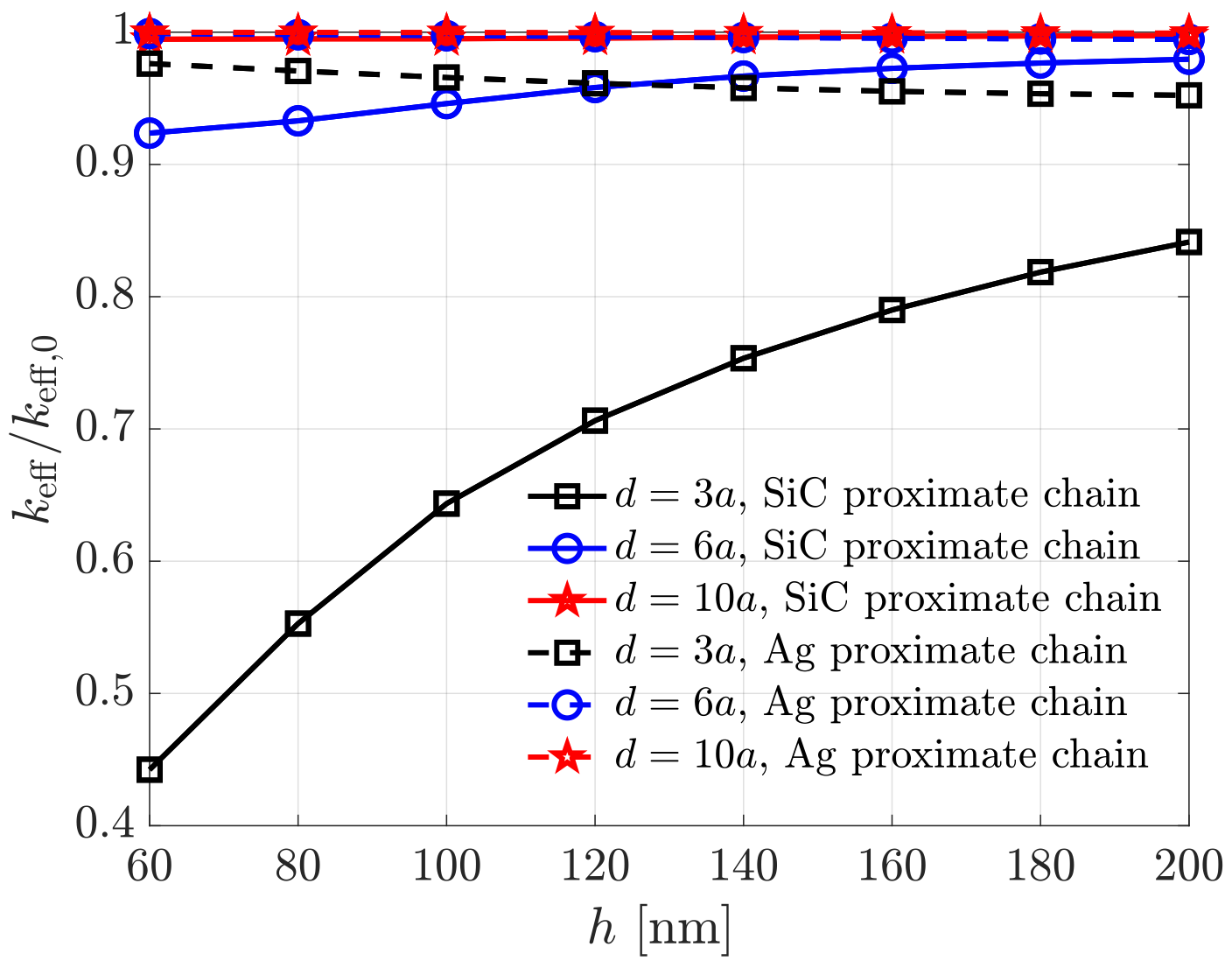}}
\caption{Dependence of $k_{\rm eff}/k_{\rm eff,0}$ on the period $h$. $k_{\rm eff,0}$ is the effective thermal conductivity of the main chain without a proximate scattering chain. The two types of proximate chains are considered, i.e., Ag and SiC nanoparticle chains.}
\label{Ag_SiC_proxi_MBI}
\end{figure*}

We then study the effect of separation ($d$) between the particle chain of interest (`main' chain) and the inserted scattering chain (`proximate' chain) on the radiative effective thermal conductivity ($k_{\rm eff}$). The main chain is made of SiC nanoparticles (radius $a=20$nm). The dependence of $k_{\rm eff}/k_{\rm eff,0}$ ($k_{\rm eff,0}$ is the effective thermal conductivity of the main chain without a proximate scattering chain)on the period $h$ for two types of proximate chains is shown in Figure~\ref{Ag_SiC_proxi_MBI}, where both SiC and Ag proximate chains are considered. $d=3a$ (60 nm), $6a$ (120 nm) and $10a$ (200 nm). Period $h$ ranges from $3a$ (60 nm) to $10a$ (200 nm).

%===== figure 7 effect of separation on ETC=======%

\begin{figure*} [htbp]
\centerline {\includegraphics[width=0.7\textwidth]{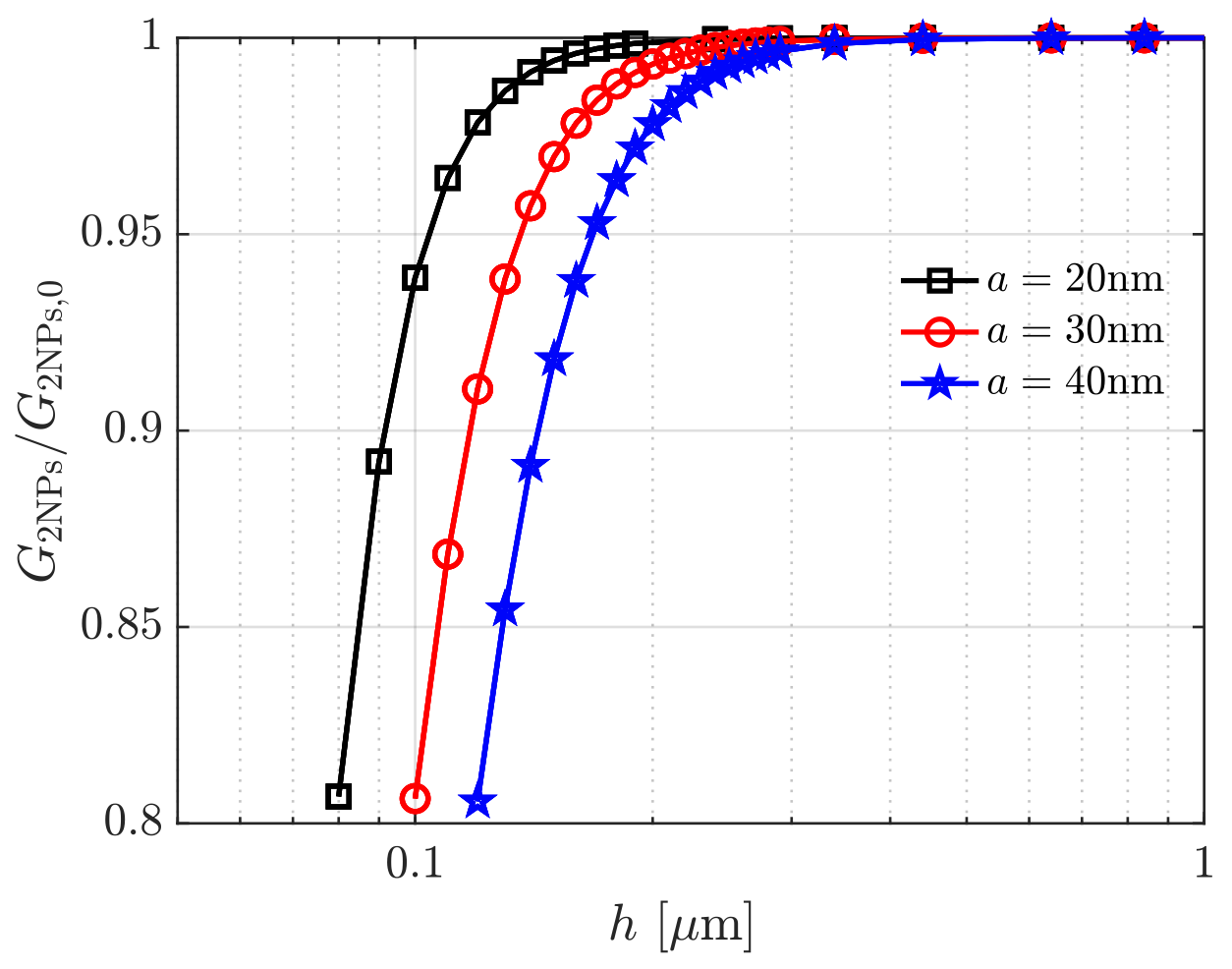}}
\caption{Dependence of $k_{\rm eff}/k_{\rm eff,0}$ on the period $h$. $k_{\rm eff,0}$ is the effective thermal conductivity of the main chain without a proximate scattering chain. The two types of proximate chains are considered, i.e., Ag and SiC nanoparticle chains.}
\label{fig_2NPs}
\end{figure*}

For a SiC proximate chain with a fixed the separation $d$, as shown in Figure~\ref{Ag_SiC_proxi_MBI}, the ratio $k_{\rm eff}/k_{\rm eff,0}$ obviously increases with the period $h$, which is due to the decreasing coupling along the chain. However, for the Ag proximate chain, the $k_{\rm eff}/k_{\rm eff,0}$ does not change too much when changing the period of the main chain, due to the weak coupling in such system mediated by the mismatch between localized surface resonances of Ag and SiC nanoparticles. In general, when increasing the separation $d$ between the inserted proximate chain and the main chain, the $k_{\rm eff}$ (with a proximate chain) is approaching to that without the proximate chain (the ratio $k_{\rm eff}/k_{\rm eff,0}$ approaches to unity). The effect of the proximate chain on $k_{\rm eff}$ decreases gradually. Because inserted proximate chain moves from the inhibiting zone to the non-MBI zone. Hence, the many-body interaction becomes less important. The transition separation where the proximate chain becomes less important can be understand from the simple case of two particle case. In our previous work \cite{Luo2020}, we proposed a ratio $G_{\rm 2NPs}/G_{\rm 2NPs,0}$ to roughly estimate the length scale $L$ to have an obvious many-body interaction. $G_{\rm 2NPs}$ is thermal conductance obtained by using the Eq.~\ref{Gij} combing with the system Green's function, which includes the multiple scattering between the two particles. $G_{\rm 2NPs,0}$ is the thermal conductance obtained by using the Eq.~\ref{Gij} combing with the Green's function in free space. We show the dependence of $G_{\rm 2NPs}/G_{\rm 2NPs,0}$ on the separation between the two particles in Figure~\ref{fig_2NPs}.

 %===== figure 10 2NPs=======%

For the particles of different size, the length scale differs from each other slightly. The length scales $L$ are 150 nm, 200 nm and 250 nm for the cases with $a=20$ nm, 30 nm and 40 nm to obtain the value 0.99 for the ratio $G_{\rm 2NPs}/G_{\rm 2NPs,0}$, where if we neglect the multiple scattering we will have about 1\% difference to the exact result. Then we can use the length scale $L=150$ nm for the particles with size $a=20$ nm to estimate the many-body interaction caused by the proximate chains in Figure~\ref{Ag_SiC_proxi_MBI}. When the separation $d=10a$(200 nm) $ > L$ (150 nm), the effect of proximate chain on effective thermal conductivity can be neglected safely. In addition, since the ratio $k_{\rm eff}/k_{\rm eff,0}$ is far from unity for the SiC proximate chains at separation $d=3a$ and approaches to unity when increasing separation $d$, hence the SiC proximate chain can be applied to reduce the heat radiation transport in the SiC main chain to reach a modulation ratio over 50\%. However, for the Ag proximate chain, the ratio $k_{\rm eff}/k_{\rm eff,0}$ is very close to unity. Thus the modulation range of heat transfer by Ag nanoparticles is limited.

\begin{figure*} [htbp]
\centerline {\includegraphics[width=0.8\textwidth]{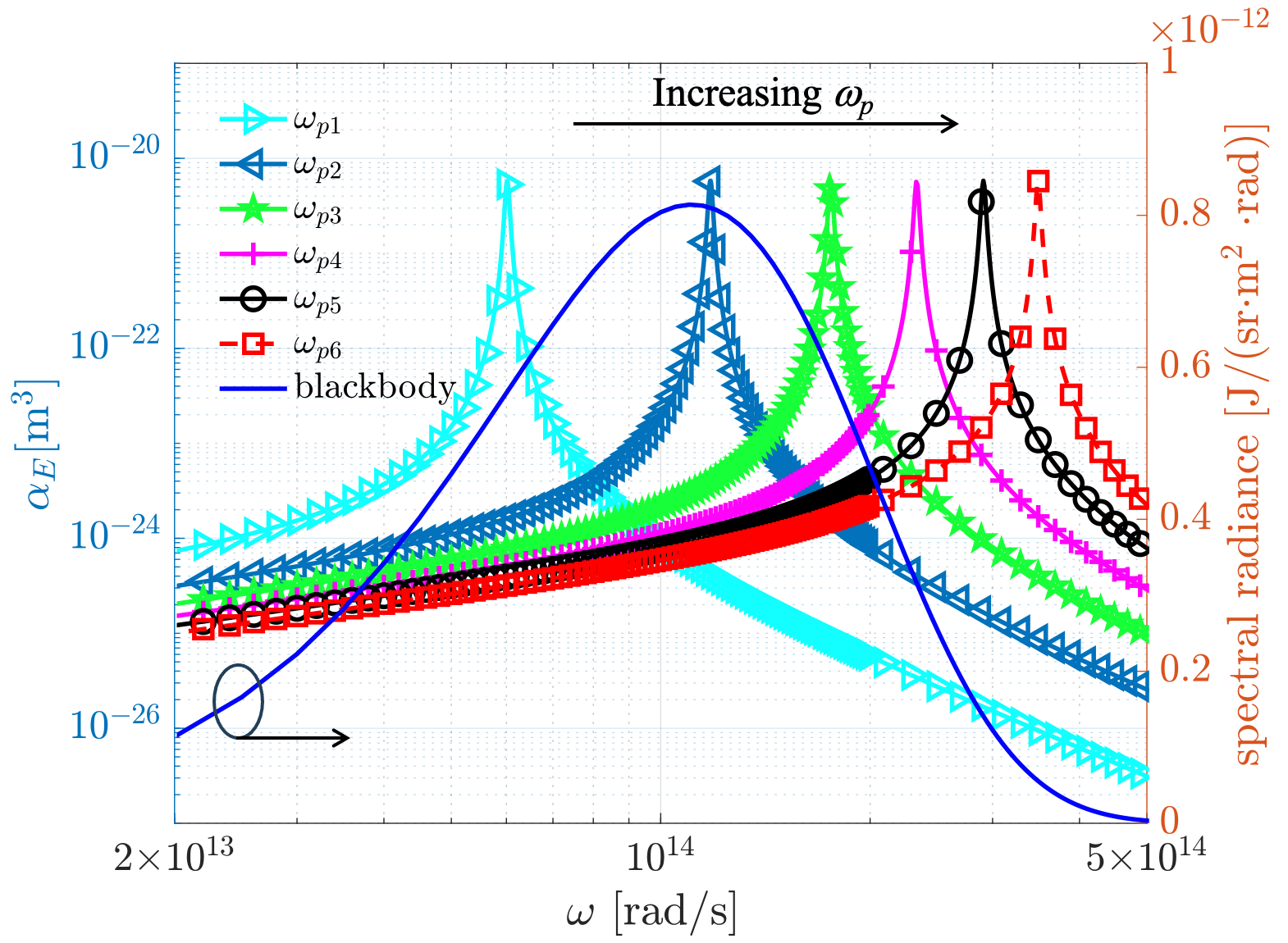}}
\caption{Polarizability spectrum of Drude particles (radius $a=20$nm). $\omega_{pj}= 10^{14}$rad$\cdot$s$^{-1}$, $2\times 10^{14}$rad$\cdot$s$^{-1}$, $3\times 10^{14}$rad$\cdot$s$^{-1}$, $4\times 10^{14}$rad$\cdot$s$^{-1}$, $5\times 10^{14}$rad$\cdot$s$^{-1}$, and $6\times 10^{14}$rad$\cdot$s$^{-1}$($j=1,~2,~3,~4,~5$ and 6). The spectral radiance of the blackbody at 300 K is added for reference.}
\label{fig_polarizability_general}
\end{figure*}

In above, we investigate the effect of many-body interaction on and the modulation of effective thermal conductivity of the network of particles composed of typical materials (e.g., dielectric SiC and metal Ag). Now we move to more general case and consider particles with the Drude-like dielectric function. Such particles support the localized surface plasmon polariton (SPP) at frequency of about $\omega_{\rm res}=\omega_p/\sqrt{3}$ \cite{Maier2007}. The SPP of Ag nanoparticle is at optical frequency in the untraviolet range, which is far from the Planckian window at room temperature. We want to investigate the impact of particles that support SPP within the Planckian range on the many-body interaction and modulation of radiative effective thermal conductivity. Following a similar the method used in Ref.~\cite{DongPrb2017}, Drude particles are introduced, of which the dielectric function is $\epsilon(\omega) = 1-\omega_p^2/(\omega^2+i\gamma\omega)$ and $\gamma=0.01\omega_p$. The $\omega_p$ is ranging from $1\times 10^{14}$rad$\cdot$s$^{-1}$ to $6\times 10^{14}$rad$\cdot$s$^{-1}$. We show the polarizability spectrum for a Drude particle (radius $a=20$ nm) in Figure~\ref{fig_polarizability_general}. When altering the $\omega_p$, the resonance moves consequently. The resonance of particle with $\omega_{p1}$ is located out of the Planckian Window. When we increase the $\omega_p$, the resonance moves in the Planckian Window. When increasing the $\omega_p$ further, the resonance moves out the Planckian window again.

%===== figure 11 polarizability of general particles=======%

\begin{figure*} [htbp]
\centerline {\includegraphics[width=0.8\textwidth]{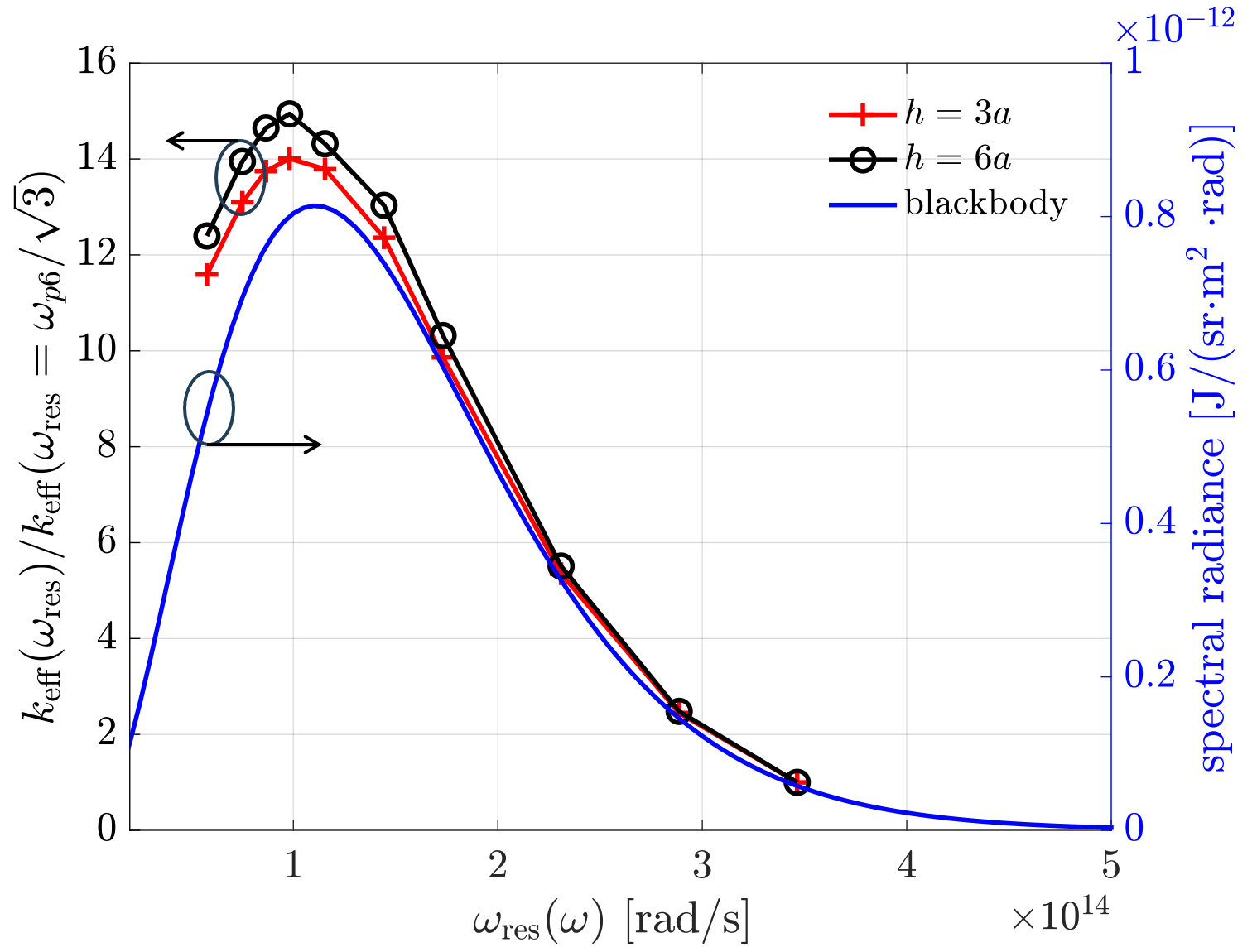}}
\caption{Dependence of $k_{\rm eff}(\omega_{\rm res})/k_{\rm eff}(\omega_{\rm res}=\omega_{p6}/\sqrt{3})$ on the $\omega_{\rm res}$. $h=3a$ and $6a$. The spectral radiance of the blackbody at 300 K is added for reference.}
\label{fig_ETC_general}
\end{figure*}

We consider the two types of ordered chains of Drude particles. The lattice spacing $h= 3a$ and $6a$, respectively. For the Drude particle, the $\omega_p$ ranges from $10^{14}$rad$\cdot$s$^{-1}$ to $6\times 10^{14}$rad$\cdot$s$^{-1}$. Thus, the corresponding angular frequency ($\omega_{\rm res}=\omega_p/\sqrt{3}$) of SPP ranges from $5.77\times 10^{13}$rad$\cdot$s$^{-1}$ to $3.46\times 10^{14}$rad$\cdot$s$^{-1}$. The radiative effective thermal conductivity of the Drude particle chain is a function of the $\omega_{\rm res}$ of the Drude particle.  We show the dependence of normalized radiative effective thermal conductivity $k_{\rm eff}(\omega_{\rm res})/k_{\rm eff}(\omega_{\rm res}=\omega_{p6}/\sqrt{3})$ on the $\omega_{\rm res}$ in Figure~\ref{fig_ETC_general}. We can observe an identical trend for the dependence of $k_{\rm eff}(\omega_{\rm res})/k_{\rm eff}(\omega_{\rm res}=\omega_{p6}/\sqrt{3})$ on $\omega_{\rm res}$ as the radiance spectrum of the blackbody. The $\omega_{\rm res}=\omega_{p6}/\sqrt{3}=3.46\times 10^{14}$rad$\cdot$s$^{-1}$ is out of the Planckian window at room temperature. The $k_{\rm eff}(\omega_{\rm res})/k_{\rm eff}(\omega_{\rm res}=\omega_{p6}/\sqrt{3})$ reaches peaks in the Planckian window for both of the two chains that $h=3a$ and $6a$. The maximums are 14 and 15, respectively. If the SPPs locate in the Planckian window, the corresponding effective thermal conductivity increases significantly. To achieve a high value of effective thermal conductivity of particle networks, one could apply the particles of the SPP in the Planckian window.

 %===== figure  keff of general particles   h=3a, 6a =======%

\begin{figure*} [htbp]
\centerline {\includegraphics[width=1.\textwidth]{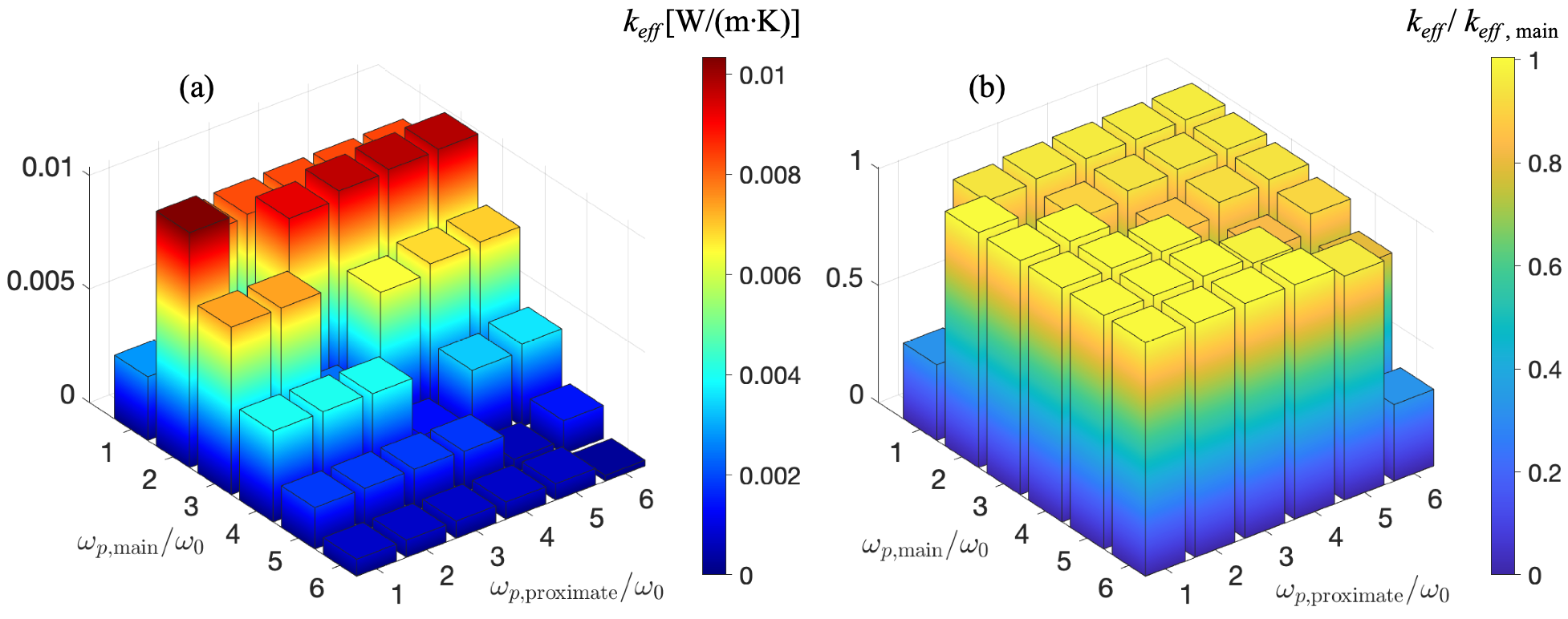}}
\caption{Dependence of the $k_{\rm eff}$ on the $\omega_{p,{\rm proximate}}/\omega_0$ and $\omega_{p,{\rm main}}/\omega_0$ ($\omega_0=10^{14}$rad$\cdot$s$^{-1}$)}
\label{fig_ETC_general_wp1_wp2_inhibition}
\end{figure*}

We then consider to insert Drude particles with various $\omega_p$ in both the inhibition zone or  enhancement zone to understand how many-body interactions affect the radiative heat transfer characteristics. When inserting Drude particles (`proximate' chain) in the inhibition zone of the `main' chain, we consider that two parallel chains are composed of two different types of Drude particles, of which the $\omega_{p}$ are denoted as $\omega_{p,~{\rm proximate}}$ and $\omega_{p,~{\rm main}}$, respectively. The dependence of the $k_{\rm eff}$ and $k_{\rm eff}/k_{\rm eff,~main}$ on the $\omega_{p,~{\rm proximate}}/\omega_0$ and $\omega_{p,~{\rm main}}/\omega_0$ is shown in Figure~\ref{fig_ETC_general_wp1_wp2_inhibition} (a) and (b), respectively. $k_{\rm eff,~main}$ is the radiative effective thermal conductivity of the `main' chain without insertion of `proximate' chain. The value of $k_{\rm eff}$ and $k_{\rm eff}/k_{\rm eff,~main}$ is denoted as the height and the color of the bar. $\omega_0=10^{14}$rad$\cdot$s$^{-1}$, the separation between the two chains $d=3a$, the lattice spacing $h=3a$, and the temperature $T=300$ K.

%===== figure  keff of general particles   h=3a, wp1~wp2 inhibition=======%

As shown in Figure~\ref{fig_ETC_general_wp1_wp2_inhibition} (a), for a fixed $\omega_{p,{\rm proximate}}$, the dependence of the $k_{\rm eff}$ on the  $\omega_{p,{\rm main}}$ shows a similar trend to that of the blackbody radiance spectrum (as shown in Figure~\ref{fig_ETC_general}). When the $\omega_{p,{\rm proximate}}$ is equal to the $\omega_{p,{\rm main}}$, the mismatch of optical properties between the particles in the main chain and the inserted particle in the proximate chain diminishes, hence the coupling between the main chain and the proximate chain maximizes and the $k_{\rm eff}$ decreases dramatically. For main chain of Drude particles with all considered values of $\omega_{p,{\rm main}}$, we can obtain a minimal effective thermal conductivity by putting a proximate chain of Drude particles of the identical  $\omega_{p,{\rm proximate}}=\omega_{p,{\rm main}}$. As shown in Figure~\ref{fig_ETC_general_wp1_wp2_inhibition} (b), the ratio $k_{\rm eff}/k_{\rm eff,~main}$ approaches to its minimal (about 0.4) when $\omega_{p,{\rm proximate}}$ equals $\omega_{p,{\rm main}}$, where many-body interaction significantly inhibits the radiative heat transfer over 60\%. When $\omega_{p,{\rm proximate}}$ differs from the $\omega_{p,{\rm main}}$, the ratio $k_{\rm eff}/k_{\rm eff,~main}$ is close to unity and the many-body interaction for radiative heat transfer becomes less important.

\begin{figure*} [htbp]
\centerline {\includegraphics[width=1.\textwidth]{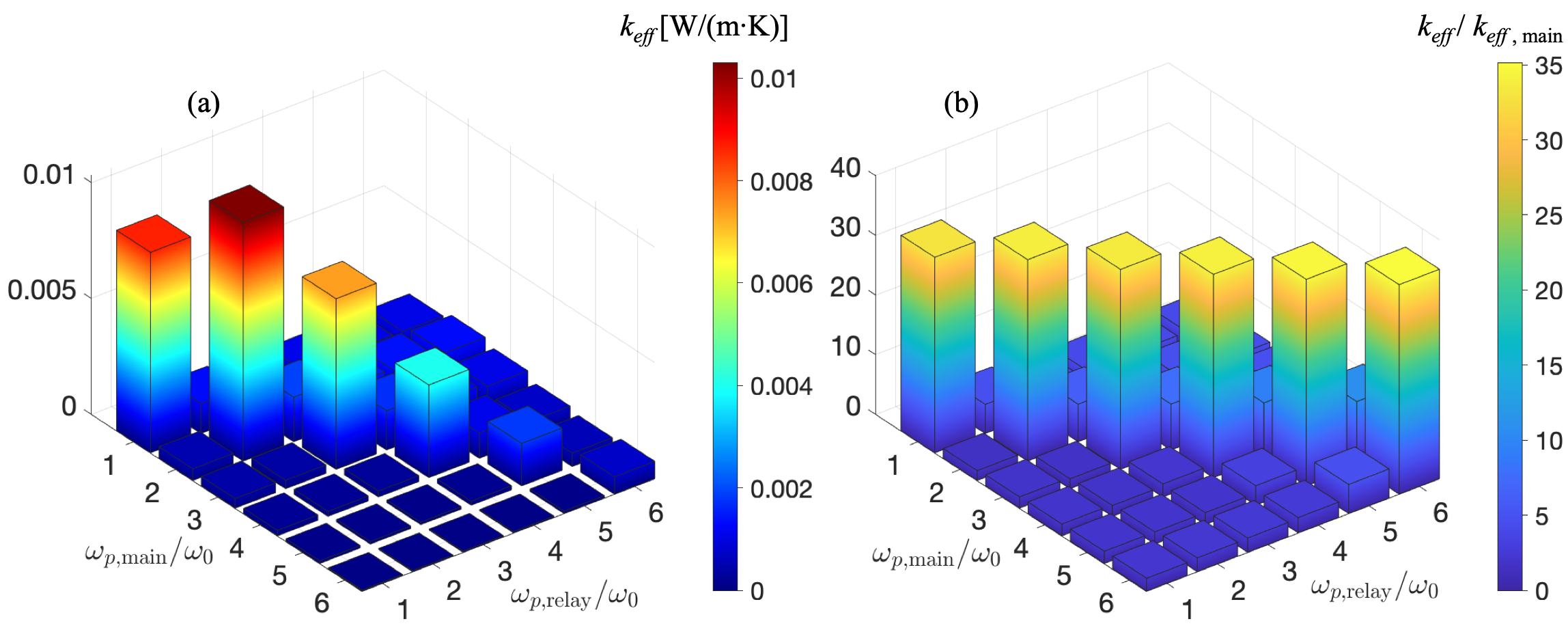}}
\caption{Dependence of the $k_{\rm eff}$ on the $\omega_{p,{\rm relay}}/\omega_0$ and $\omega_{p,{\rm main}}/\omega_0$ ($\omega_0=10^{14}$rad$\cdot$s$^{-1}$)}
\label{fig_ETC_general_wp1_wp2_enhancement}
\end{figure*}

We then consider inserting Drude particles in the enhancement zone to see many-body interactions for the radiative effective thermal conductivity. We add particles (`relay') in arbitrary two adjacent (`main') particles in case 1, of which the positions are shown as shading of Case 1 of Figure ~\ref{ETC_geo} to construct a more dense particle network, i.e., Case 2. We focus on the $k_{\rm eff}$ for the Case 2 and consider two different types of Drude particles (the $\omega_{p}$ are denoted as $\omega_{p,{\rm relay}}$ and $\omega_{p,{\rm main}}$, respectively). The dependence of the $k_{\rm eff}$ and the ratio $k_{\rm eff}/k_{\rm eff,~main}$ on the plane of $\omega_{p,{\rm relay}}/\omega_0$ and $\omega_{p,{\rm main}}/\omega_0$ is shown in Figure~\ref{fig_ETC_general_wp1_wp2_enhancement} (a) and (b), respectively. The value of $k_{\rm eff}$ and $k_{\rm eff}/k_{\rm eff,~main}$  is also denoted as the height and the color of the bar. $\omega_0=10^{14}$rad$\cdot$s$^{-1}$, the lattice spacing $h_2=3a$ and the temperature $T=300$ K.

%===== figure  keff of general particles   h=3a, wp1~wp2 enhancement =======%

As shown in Figure~\ref{fig_ETC_general_wp1_wp2_enhancement} (a), for a fixed $\omega_{p,{\rm main}}$, the $k_{\rm eff}$  increases much significantly (about 35 times of magnitude of that for the case without inserted particles, as shown in Figure~\ref{fig_ETC_general_wp1_wp2_enhancement} (b)) when the $\omega_{p,{\rm relay}}$ of the inserted particles is equal to $\omega_{p,{\rm main}}$ of the main particles, because of the strong coupling induced by the inserted particles. This statement holds true for all considered $\omega_{p,{\rm main}}$, regardless of whether they are located within the Planckian window or not. That is to say the inserted particles can work as a relay helping for radiative heat transfer when the optical properties of inserted particles match well with that of the particles in the `main' chain. For a fixed value of $\omega_{p,{\rm relay}}$, the $k_{\rm eff}$ dependence on $\omega_{p,{\rm main}}$ shows a trend similar to that of the blackbody radiance spectrum (as indicated in Figure~\ref{fig_ETC_general}), except for the significant increase in $k_{\rm eff}$ when $\omega_{p,{\rm main}}$ equals $\omega_{p,{\rm relay}}$. In addition, if we focus on the cases $\omega_{p,{\rm main}}$ equals $\omega_{p,{\rm relay}}$, the $k_{\rm eff}$ dependence on $\omega_{p}$ ($\omega_{p,{\rm main}}$=$\omega_{p,{\rm relay}}$) also shows a trend similar to that of the blackbody radiance spectrum (as indicated in Figure~\ref{fig_ETC_general}).

{To have  a significant many-body interaction regardless of enhancement or the inhibition for radiative heat transfer, it is necessary to introduce particles that have resonances well-matched with those of the particles of interest. It is worthwhile to mentioning that once we can adjust the resonances of particles, we can control the many-body interactions for radiative heat transfer as desired, which holds substantial importance for experimental investigations. Finally, we want to emphasize that the resonance can be adjusted by changing the shape of particles, in situ by merely changing the chemical potential if particles are covered with graphene (which can be made e.g. via a gate voltage applied to the structure), and by applying external magnetic field to the magneto-optical particles, to name a few.}

\section{Conclusion}
Effect of many-body interactions on near-field radiative heat transfer in nanoparticle networks is analyzed by means of the many-body radiative heat transfer theory at particle scale and the normal-diffusion radiative heat transfer theory at continuum scale. An influencing factor $\psi$ is defined to numerically figure out the different many-body interaction regimes. The space near the two nanoparticles can be divided into four zones, non-MBI zone, enhancement zone, inhibition zone and forbidden zone, respectively. Obvious boarder of different regimes of many-body interaction on NFRHT is distinguished. Enhancement zone is relatively smaller than the inhibition zone. In the realistic nanoparticle network, a lot of particles will inevitably lie in the inhibition zone, which accounts for the inhibition effect of many-body interaction on NFRHT (commonly seen in literature \cite{Dong2017JQ,Chen2018JQ,Luo2019JQ,Luo2019}). The many-body interaction for radiative heat transfer caused by insertion of a third particle is confirmed by analysis on the radiative thermal energy. Insertion of a third particle will significantly enhance the radiative thermal energy by several orders of magnitude when the third particle lies in the center of the two particles (corresponding to the enhancement zone) and will slightly inhibit the radiative thermal energy when the third particle lies in proximity of the receiver particle (corresponding to the inhibition zone).

As for effect of MBI on radiative heat diffusion characteristics (characterized by the effective thermal conductivity), we start the analysis from the typical dielectric SiC and metal Ag to a kind of generalized Drude particles. We find that a good match between the resonance and the Planckian window can guarantee a high effective thermal conductivity. Inserting particles in the enhancement zone can significantly enhance the radiative thermal conductivity by over 30 times of magnitude, which is due to that the added particles in the enhancement zone work as relays for near-field photon tunneling and radiative heat transfer. Inserting proximate particles in the inhibition zone can reduce the radiative effective thermal conductivity by over 55\% due to MBI. To achieve a notable many-body interaction for radiative heat transfer (enhancement or inhibition), it is necessary to introduce particles that have resonances well-matched with those of the particles of interest, irrespective of their match with the Planckian window. By arranging the system's particles (structures and optical properties), many-body interactions can be used to control radiative heat diffusion over a wide range, from inhibition to amplification. This work may help for the understanding of thermal radiation in nanoparticle networks.

\section*{Acknowledgements} 
The support of this work by the National Natural Science Foundation of China (No. 51976045 and 52206081) is gratefully acknowledged. The supports from the China Postdoctoral Science Foundation(2021M700991) and China Scholarship Council (No.201906120208) are also acknowledged.

\addcontentsline{toc}{section}{Acknowledgements}

\bibliographystyle{Bibliography_Style}

\bibliography{MBI_RAD23_SI}

\end{document}